\documentclass[aps,pre,twocolumn,superscriptaddress,showpacs]{revtex4}
\usepackage{bm}
\usepackage{amsmath}
\usepackage{amssymb}
\usepackage{graphicx}

\begin{document}

\title{Experimental evidence of accelerated energy transfer in turbulence}

\author{R. Labb\'e}
\affiliation{Laboratorio de Turbulencia, Departamento de F\'isica,
Facultad de Ciencia, Universidad de Santiago de Chile. Casilla
307, Correo 2, Santiago, Chile}
\author{C. Baudet}
\affiliation{Laboratoire des Ecoulements geophysiques et
Industriels, UMR 5519 CNRS/UJF/INPG.\\ 1025, Rue de la Piscine,
38041 Grenoble, France}
\author{G. Bustamante}
\affiliation{Laboratorio de Turbulencia, Departamento de F\'isica,
Facultad de Ciencia, Universidad de Santiago de Chile. Casilla
307, Correo 2, Santiago, Chile}

\date{21 June 2005; Revised 09 June 2006}

\begin{abstract}
We investigate the vorticity dynamics in a turbulent
vortex using scattering of acoustic waves. Two ultrasonic
beams are adjusted to probe simultaneously two spatial scales in a
given volume of the flow, thus allowing a dual channel recording
of the dynamics of coherent vorticity structures. Our results show
that this allows to measure the average energy transfer time
between different spatial length scales, and that such transfer
goes faster at smaller scales.
\end{abstract}



\maketitle

Much of the investigation in turbulence has been devoted to the
understanding of the mechanisms underlying the energy transfer
through the turbulent cascade, a concept introduced by Richardson in
1922 \cite{Frisch}. A good deal of theoretical and numerical work
has been dedicated to study shell models, which include the
essential features expected in a turbulent cascade, but without some
of the complications inherent to the Navier-Stokes equations. Though
these models retain some of the dynamics of the motion equations,
they handle the velocity field as a scalar. Thus the price payed is
a complete loss of flow geometry \cite{Bife}. Another approach is
the study of random multiplicative cascade models, which adequately
mimic the statistics of the local energy dissipation rate $\epsilon$
at different flow scales, and the intermittency of small scales
observed in a large number of experimental results. Although much
progress was done, many issues related to the statistics of energy
transfer, energy dissipation and small scale intermittency in
turbulent flows still remain without satisfactory answers. In
particular, the dynamics of the energy transfer through the
turbulent cascade has been only recently addressed in some
theoretical works \cite{Leve,Benz}, but this aspect of turbulent
flows is still lacking experimental studies specifically addressing
it.

In this note we report an experimental result, obtained for the case
of a vortex embedded in a turbulent flow produced in air by two
coaxial centrifugal fans, each facing the other and rotating at
fixed angular velocities. This so-called von K\'arm\'an flow has
been used in a number of experiments in turbulence \cite{All}. When
the fans rotate in the same direction with identical speeds, a
strong vortex is produced between them \cite{LPF}. The advantage of
this configuration is that in a small volume we obtain an intense
and well-defined coherent vorticity structure, surrounded by a fully
turbulent background. Details of this setup are given in a previous
work \cite{LP}. The parameters used in this case are: diameter of
fans: D = 30~cm, height of vanes: $h = 2.2$~cm, distance between
disks: $H = 30$~cm, rotation speed: $f = 30$~Hz, and diameter of
central holes: $d = 2.5$~cm. The fans were driven by two DC motors,
powered by independent constant voltage sources allowing to keep the
disk rotation within $1$\% of the desired speed. Local measurements
of the airflow speed were performed with a Dantec Streamline 90N10
frame housing a Constant Temperature Anemometer which drives a
$5\thinspace\mu$m diameter and $2$~mm length platinum wire probe. A
Streamline 90H02 Flow Unit was used to calibrate it. To produce and
detect the sound waves, four circular Sell type electrostatic
transducers \cite{Ank} were used, each having an active surface
diameter of 14~cm. Two B\&K model 2713 power amplifiers drove the
emitters. The receivers used custom made charge amplifiers, whose
output were connected to two Stanford Research model SR830 lock-in
amplifiers (LIA) through $2^{\textrm {nd}}$ order high-pass RC
filters having cut-off frequencies of 7.5~kHz. The sine-wave output
of each LIA internal generator was used as signal source for the
corresponding emitter. The LIA output signals, comprising in-phase
and quadrature components, were low-pass filtered by an IOTech
Filter-488 $8^{\textrm {th}}$ order elliptic filter bank, using a
cutoff frequency of 4~kHz. The resulting signal was digitized at
12.5~kS/s with a 16-bit resolution National Instruments AT-MIO-16X
multifunction board, installed into a personal computer. A total of
$2^{20}$ points per signal channel was collected in each of 77 data
sets.

\begin{figure}[t]
\centering \vspace{-0.2cm} \hspace{-0.2 cm}
\includegraphics[width=.42\textwidth]{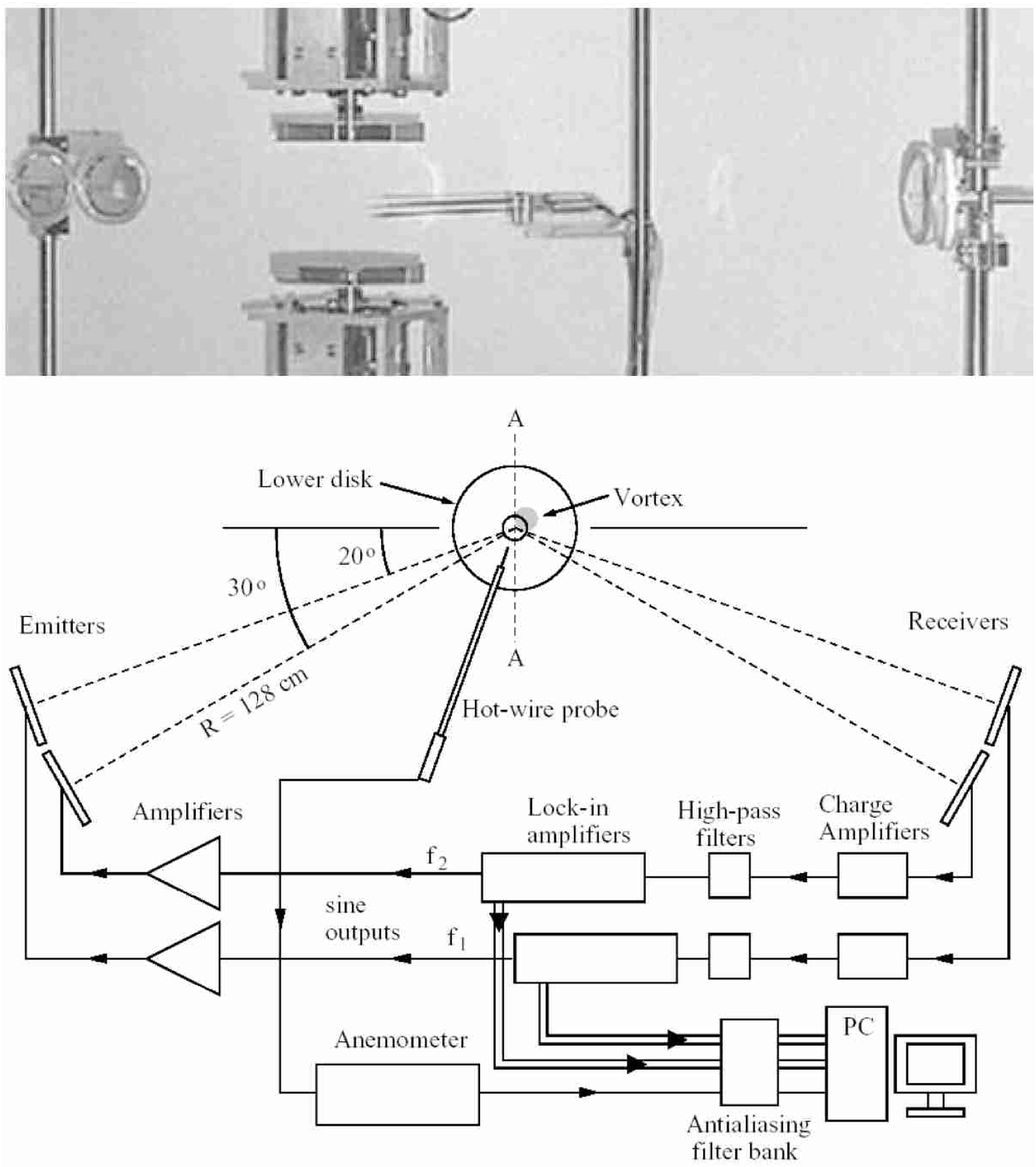}
\caption{{Top: Partial view of the experimental setup. Bottom: a
schematic drawing of the experimental setup  (see text).}}
\label{one}
\end{figure}

Fig.~1 (top) shows the device that produced the flow, along with the
four circular ultrasonic transducers, and the hot-wire probe. The
transducers were arranged to perform measurements using an
interferometric configuration, described in detail in a previous
work \cite{Baud}. The scattering angles were $\theta_{1}=40^{\circ}$
and $\theta_{2}=60^{\circ}$. The magnitude of the scattering
wave-vectors,
$q_{i}^{scatt}=4\pi\nu_{0,i}\sin(\theta_{i}^{scatt}/2)/c=|\bm{k'}-\bm{k}|,
i=1,2$, were adjusted independently for each channel through their
corresponding frequencies $\nu_{0,i}$. We acquired three sets of
data, setting $\nu_{0,1}=20$ kHz, $40$ kHz, and $60$ kHz; while
$\nu_{0,2}$ took some 25 values in the interval $[16, 90]$ kHz in
each case. Notice that each probed scattering wave-vector, related
to the $z$\thinspace -component of the vorticity distribution, can
be set arbitrarily close (even identical) to the other.  Fig.~1
(bottom) displays a schematic upper view of the lower disk, hot-wire
probe and ultrasonic transducers, along with distances and angles.
This arrangement is symmetric with respect to the vertical plane AA.
Also displayed is a schematic diagram of the measuring chain. Fig.~2
(top) is a sketch of the volume probed by the ultrasonic beams,
showing the relationships between scattering vectors and
wave-vectors in the case of exact inter-channel tuning (same
scattering wave-vector). As mentioned in previous works
\cite{LPF,LP}, the vortex performs a slow precession motion around
the axis of rotation of the fans. Even so, it remains inside the
measurement volume, as shown in fig.~2. A horizontal cut of the
measurement volume is displayed as a shaded rhombus, the heavier
shaded circle indicating the position of the vortex core. Air speed
measurements with the probe wire aligned vertically, and located at
about 2~cm of the disks's rotation axis and some 10~cm above the
lower disk, give the sample record displayed in fig.~2 (bottom). The
nearly periodic precession of the vortex is visible as a pattern
with sharp dips ---sometimes enlarged or doubled by fluctuations due
to the turbulent background, each time the center of the vortex is
close or coincides with the hot wire location. These events are
signaled by arrows in fig.~2, the time interval between single
events being about $1$~s. Accordingly, the spectrum displayed in
fig.~3 (top) has a rather wide peak located at $f_{0}\approx1$~Hz.
The Taylor hypothesis is not met by this flow, but this spectrum
gives an idea of the energy contents at different scales. At higher
frequencies (smaller scales), it meets only approximately the
$k^{-5/3}$ K41 law (represented by the straight line) in the
inertial range, due to the presence of the vortex in the bulk of the
flow. In fact, we observe three regions, A, B, and C, where the
spectrum looks self-similar, but with different slopes. As we will
see later, this could be related to the influence of the flow
anisotropy on the dynamics of the energy transfer through the
turbulent cascade. The acoustic signals are delivered by the LIAs as
low frequency complex voltages, which are images of the spatial
Fourier modes $\Omega_{z}(q_{j},t)$ of the flow vorticity, probed at
well-defined wave vectors $q_{j}=4\pi\nu_{0,j}\sin(\theta_{j}/2)/c$,
$j=1,2$. Typical spectra of these signals are displayed in fig.~3
(bottom). The wide central peak, due to the intense coherent
vorticity  of the vortex, lies between two side bands produced by
the background flow. There is a close correspondence between the low
frequency region of the velocity spectrum (up to $50$~Hz) and the
central region of the vorticity spectrum. The same is true for the
high frequency region ($>200$~Hz) of the spectra, both exhibiting a
slowly decaying roll-off, with a slope close to $-5/3$ related to
the background turbulence of the flow.

\begin{figure}[t]
\centering \vspace{-0.2cm} \hspace{-0.5 cm}
\includegraphics[width=.45\textwidth]{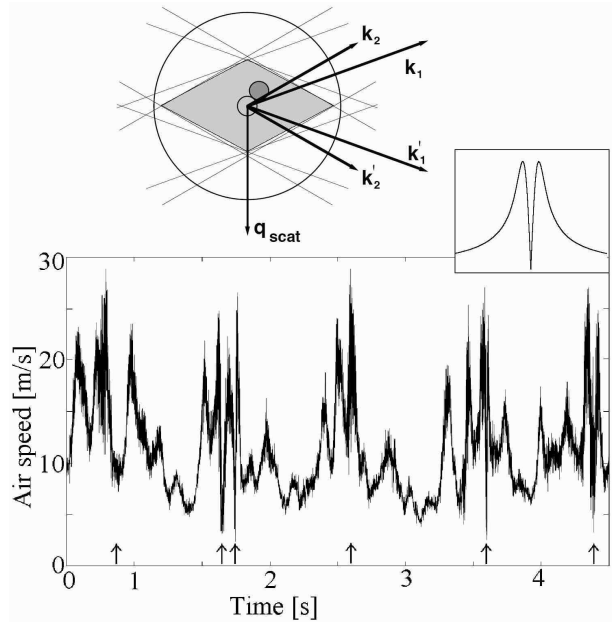}
\caption{{Top: Diagram showing the relationship between
wave-vectors, the volume probed by the ultrasonic beams (shaded
region) and the size of the disks. The dark circle represents a cut
of the vortex core. Bottom: A sample of the anemometer signal.
Arrows signal events of minimal air speed, the vortex core being
close to the hot wire probe there. The inset displays the velocity
profile of a non turbulent vortex, which helps in recognizing the
vortex core in the turbulent signal.}} \label{two}
\end{figure}

\begin{figure}[t]
\centering \vspace{-0.2cm} \hspace{-0.5 cm}
\includegraphics[width=.45\textwidth]{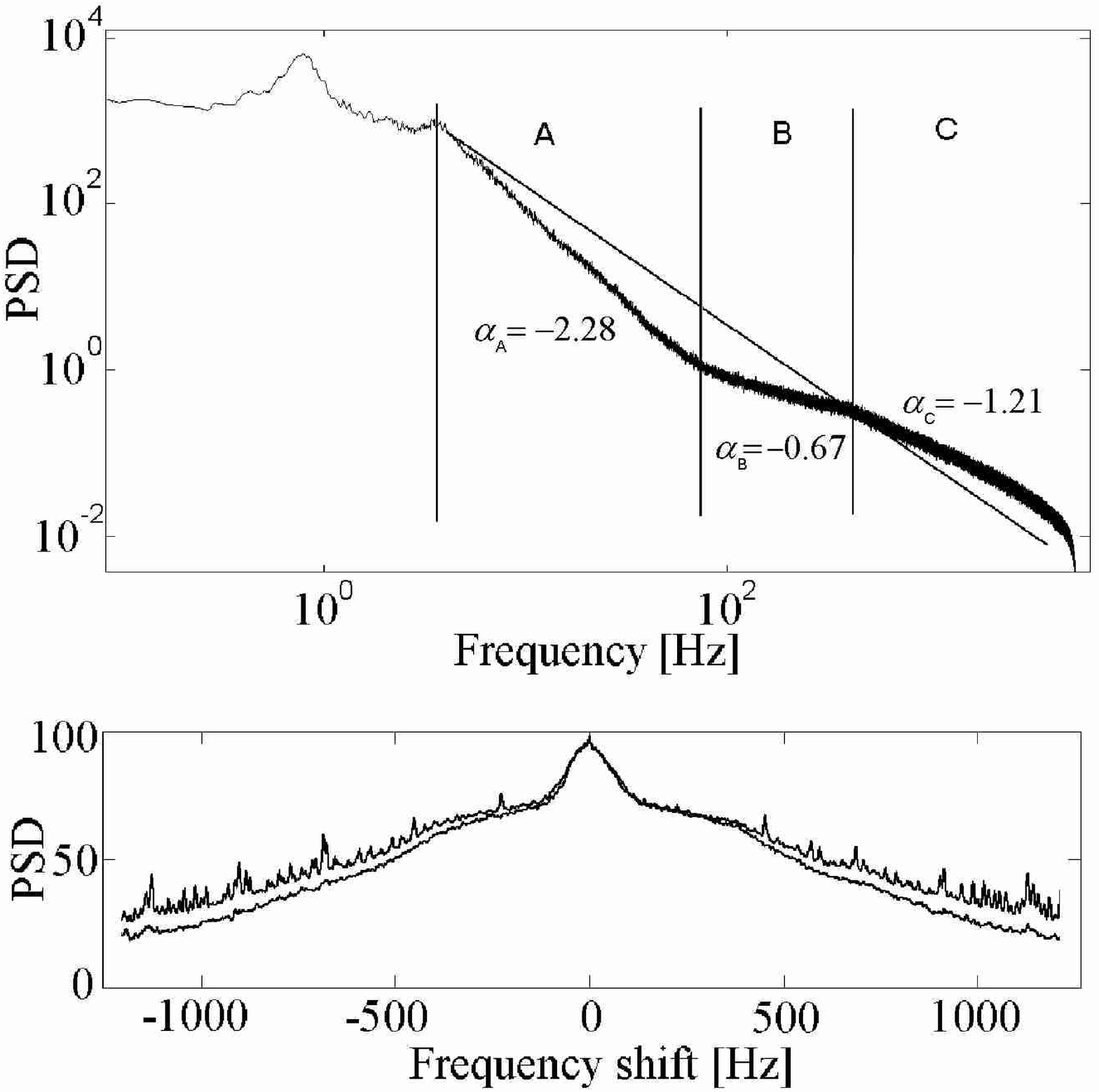}
\caption{{Top: Power spectral density of the signal displayed in
fig. 2. Three almost linear regions are labeled A, B, and C (see
text), with slopes $\alpha_{A,B,C}$. The straight line represents
the Kolmogorov spectrum, with slope $\alpha_{K}=-1.66$. Bottom:
Individual spectra of the signals delivered by the lock-in
amplifiers. The spectrum corresponding to the scattering angle of 60
degrees decreases slightly faster with the frequency shift.
Scattered waves having the same central frequency of the incident
beam have no frequency shift.}} \label{three}
\end{figure}

Now we address the separation of the coherent structure and the
background turbulent flow. Following ref. \cite{Baud}, we compute
the coherence between two spatial Fourier modes of the vorticity,
$\Omega_{z}(q_{1},t)$ and $\Omega_{z}(q_{2},t)$. The coherence
function, $C(q_1,q_2,\nu)=\langle
\Omega_{z}(q_{1},\nu)\Omega^{\ast}_{z}(q_{2},\nu) \rangle/\sqrt
{\langle|\Omega_{z}(q_{1},\nu)|^{2}\rangle\langle|\Omega_{z}(q_{2},\nu)|^{2}\rangle}$,
where $\nu$ is a frequency shift associated to the Fourier transform
in time of $\Omega_{z}(q_{i},t)$, is a statistical estimator that
allows to evidence phase stationarity (coherence) between distinct
Fourier modes. Fig.~4 (top) is a plot of the coherence between two
scattering channels tuned to probe the same scale of the vorticity
distribution $(q_{1}=q_{2})$. A central peak with a high level of
coherence ($0.9$, being $1$ the maximum expected value; the value of
$0.9$ should be ascribed to incoherent noise) reveals that both
channels are detecting the same coherent structure. This was
verified ``turning off'' the vortex by rotating the disks in
opposite directions at the same angular speed. In this case only a
turbulent shear layer remains, and the plot of the coherence
function displays only the side bands. By stacking plots of
coherence for a sequence of values of $q_1$, it is possible to build
surfaces of coherence relating events at different flow scales, as
functions of $\Delta q=q_2-q_1$ and the frequency shift $\nu$, as
shown in fig.~4. We can interpret them in as eddies at given scales
drawing its energy from decaying structures at larger scales, in the
frame of the Richardson's turbulent cascade. As these surfaces
represent a time average of the whole story, we perform a time
cross-correlation analysis to confirm this interpretation. This is
done by computing the cross-correlation $\chi(q_1,\Delta q,\Delta
t)=\langle |\Omega_z (q_1,t)||\Omega_z (q_1 +\Delta q,t+\Delta t)|
\rangle/\sqrt {\langle|\Omega_z (q_{1},t)|^2 \rangle\langle|\Omega_z
(q_1 +\Delta q,t+\Delta t)|^2 \rangle}$, were
$|\Omega_{z}(q_{i},t)|$ is a measure of the strength of a structure
with vorticity $\Omega_{z}(q_{i},t)$. The function $\chi(q_1,\Delta
q,\Delta t)$ will have a peak at $\Delta t = \tau$ ($\tau = $ time
lag), indicating the presence of a vorticity pattern at the scale
$l_{1}=2\pi/q_{1}$ in time $t$ that appears at the scale
$l_{2}=2\pi/q_{2}$ a time $t+\tau$ later. The peak amplitude is a
measure of how well the pattern is preserved from one scale to the
other, a value of one meaning a perfect preservation. Fig.~5 shows
three straight lines resulting from fits to the time lags calculated
from such cross-correlation analysis, each corresponding to one of
the coherence surfaces shown in fig.~4. The time lags found here are
consistent with those given in reference \cite{Baud}. We have $\tau
\sim 1$~ms for $\Delta q=8$~cm$^{-1}$, and the transfer rates of
fig.~7 (c) in reference \cite{Baud} are similar \cite{note}.

\begin{figure}[t]
\centering \vspace{-0.3cm} \hspace{0.5 cm}
\includegraphics[width=.45\textwidth]{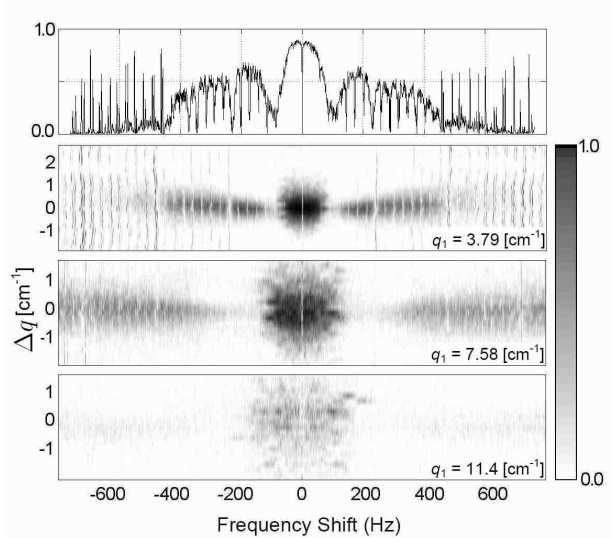}
\caption{{Top: coherence plot of signals detected simultaneously by
both receivers. The wide peak centered at zero frequency shift is
due to the vortex. Bottom: Three coherence surfaces are shown,
obtained by detuning one of the emitters by a frequency
corresponding to a shift $\Delta q$ in the wave number.}}
\label{four}
\end{figure}

Our main result is the experimental observation of acceleration in
the energy transfer in turbulence, evidenced by the systematic
change in the trend of the data in fig. 5 when the magnitude of the
reference wave-vector $q_{1}$ is changed. The slopes of the fitted
straight lines decrease with increasing $q_{1}$. The scattering of
the points is an indication of the intrinsic variability of the
energy transfer time through the turbulent cascade ($R^{2}\geq 0.6$
in all cases). The following argument helps to understand the trend
change: the time lag measured between the wave numbers $q_{1}$ and
$q_{2}=q_{1}+\Delta q$ can be seen as the average transfer time of
the energy contained in a scale corresponding to $q_{1}$ to a
smaller scale corresponding to $q_{2}$. Thus, the slope $\delta
\tau/\delta q$ (we drop $\Delta$'s from now on) of the straight
lines can be related to the inverse of the average transfer rate of
energy in the $k$\thinspace -space, in a neighborhood of the
reference scale corresponding to the wave number $q_{1}$
\cite{Fauve}. This decrease in slope with increasing $q_{1}$ reveals
that the average transfer time of the energy decreases when the
involved scales are smaller. In other words, taking $\delta q/\delta
\tau$ as the average transit speed (in the $k$\thinspace -space) of
structures starting at the wave number $q_{1}$ and ending at the
wave number $q_{2}$, we conclude that such speed is larger at larger
$q$'s (or smaller scales). This is consistent with a remark made in
a previous numerical work \cite{Leve}. A rather crude explanation of
this result follows: if $E(q)$ is the spectral density of energy per
unit mass ---averaged in time, and we think of $\delta q/\delta
\tau$ as being an ordinary derivative, then $E(q)\delta q/\delta
\tau$ is the average energy flux density in $k$\thinspace -space.
But this is the average dissipation rate per unit mass $\epsilon$.
Thus, $E(q)\delta q/\delta \tau=\epsilon$, and $\delta \tau/\delta
q=E(q)/\epsilon$. As $E(q)$ decreases with increasing $q$, so does
$\delta \tau/\delta q$. Notice that in this picture, the energy
attached to a given eddy is transported towards larger wave numbers
merely by reduction of its spatial scale (in directions
perpendicular to its vorticity) due to stretching by the velocity
field of larger eddies.

\begin{figure}[t]
\centering \vspace{0.0cm} \hspace{-0.5 cm}
\includegraphics[width=.45\textwidth]{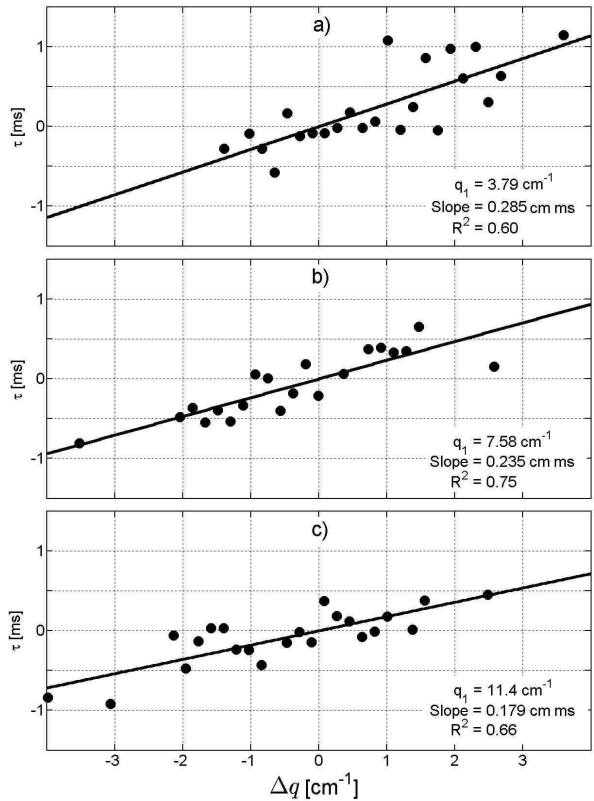}
\caption{{Time lags corresponding to the coherence surfaces of fig.
4. a), b), and c) correspond to
$q_{1}=3.79,~7.58,$~and~$11.4$~cm$^{-1}$, respectively. The straight
lines are linear fits to the measured points, with correlation
coefficient $R^{2}=0.60~(0.75, 0.66)$. Note that the slopes decrease
with increasing $q_{1}$.}} \label{five}
\end{figure}

We can illustrate our view in the frame of K41 theory \cite{Frisch}:
we have $E(q)=C_{\text{K}}\epsilon^{2/3}q^{-5/3}$, where
$C_{\text{K}}$ is the Kolmogorov constant. Thus, for homogeneous and
isotropic turbulence, $\delta \tau/\delta
q=C_{\text{K}}\epsilon^{-1/3}q^{-5/3}$, which is of course a
decreasing quantity. By rearranging terms, we have $q^{-5/3}\delta
q=(\epsilon^{1/3}/C_{\text{K}})\delta t$, which by integration
yields $q(t)=(q_{0}^{-2/3}-\frac{2}
{3}\epsilon^{1/3}t/C_{\text{K}})^{-3/2}$. This is the itinerary of
the average magnitude $q$ in a homogeneous and isotropic turbulent
flow in which the mean dissipation rate per unit mass is $\epsilon$.
The scale $l_0=2 \pi/q_0$, corresponding to $t=0$, can take any
value within the inertial range. Notice that $q(t)$ diverges in a
finite time $t_{\text{K}}^{0}=\frac{3}{2}C_{\text{K}}\thinspace
\epsilon^{-1/3}q_{0}^{-2/3}$, corresponding to the average time
taken by the energy at the scale $l_{0}$ to cascade down to
$l=0^{+}$ ($q\rightarrow \infty$), in a fluid with vanishing
viscosity ---in the frame of K41 theory. Putting
$C_{\text{K}}\approx 0.5$, $\epsilon\sim 10^{2}$~W/kg, estimated
from the power injected to the flow ($\approx 70$~W) and the volume
of moving air ($\sim 0.5$~m$^{3}$) and, for instance, $q_{0}\approx
3\times 10^{2}$ m$^{-1}$, corresponding to the largest scale
measured with our acoustic method, we get $t_{\text{K}}^{0}\sim
4$~ms. This is on the order of characteristic times in fig.~5,
showing that transfer times in homogeneous turbulence do not differ
too much from those in our flow. Thus, our reasoning shows that the
trend observed in fig.~5 can be understood in terms of ``classical''
homogeneous and isotropic turbulence. In non-homogenous,
non-isotropic turbulence, as in the present experiment, the behavior
of $\delta \tau/\delta q$ can be estimated from the experimental
spectrum shown in fig.~3, where the Kolmogorov spectrum is plotted
as a straight line, for comparison. In the large scales region (A),
$\delta \tau/\delta q$ decreases faster than in K41 turbulence, the
cascade acceleration being larger. In the small scales region (C),
$\delta \tau/\delta q$ tends to reach the Kolmogorov slope, but the
acceleration is slightly smaller than in K41. Region (B), in which
the cascade acceleration reaches its minimum, is a transition zone
from a non-isotropic flow at large scales, where vorticity is mainly
parallel to the $z$\thinspace -axis, to a more isotropic small-scale
flow. In all three regions the spectrum seems to be self-similar,
with different scaling exponents. Of course, being this one a first
experimental result revealing the accelerated nature of energy
transfer through the turbulent cascade, more work will be necessary
to state our conclusions on more solid ground. We recall that we are
measuring only the $z$\thinspace -component of the vorticity and the
modulus of the velocity component normal to the $z$\thinspace -axis.
It should be necessary to simultaneously measure at least one more
component of the vorticity. Additionally, the conditions of validity
for the Taylor hypothesis are not meet in these hot-wire
measurements ---a flying probe would be much better to study this
flow. At any rate, in the large-scale range the vorticity in our
flow is mainly parallel to the $z$\thinspace -axis. Thus, eddies in
the large scale region have mostly parallel vorticities, and the
vortex stretching is largely inhibited there, slowing down the
energy transfer. At smaller scales, where a more isotropic flow
exists, vortex stretching is reestablished, giving faster energy
flux rates. Thus, in-between the acceleration should be greater than
in the isotropic case. This could explain in part our results. To
conclude, we remark that our reasoning suggest that cascade
acceleration may be a necessary but not sufficient condition for
small scale intermittency.

Financial support for this work was provided by FONDECYT, under
projects \#1990169, \#7990057 and \#1040291.


\begin{thebibliography}{99}
\bibitem{note} We have not dropped $2\pi$ factors here in the magnitude
of scattering vectors, which explain the difference in $\Delta q$
scales in the aforementioned figures.
\end{thebibliography}
\end{document}